# Bonding and Electronic Nature of the Anionic Framework in $LaPd_3S_4$

Tanya Berry,[1,2*] Michael Nicklas,[3] Qun Yang,[3] Walter Schnelle,[3] Rafał Wawrzyńczak,[3] Tobias Förster,[4] Johannes Gooth,[3] Claudia Felser,[3] Tyrel M. McQueen[1,2,5*]

[1]Department of Chemistry, The Johns Hopkins University, Baltimore, Maryland 21218, USA
[2]Institute for Quantum Matter, William H. Miller III Department of Physics and Astronomy, The Johns Hopkins University, Baltimore, Maryland 21218, USA
[3]Max Planck Institute for Chemical Physics of Solids, Nöthnitzer Straße 40, 01187 Dresden, Germany
[4]Hochfeld-Magnetlabor Dresden (HLD-EMFL) and Würzburg-Dresden Cluster of Excellence ct.qmat, Helmholtz-Zentrum Dresden-Rossendorf, 01328 Dresden, Germany
[5]Department of Materials Science and Engineering, The Johns Hopkins University, Baltimore, Maryland 21218, USA

*Email: tberry@ucdavis.edu and mcqueen@jhu.edu

**Abstract** Double Dirac materials are a topological phase of matter in which a non-symmorphic symmetry enforces greater electronic degeneracy than normally expected – up to eightfold. The cubic palladium bronzes $NaPd_3O_4$ and $LaPd_3S_4$ are built of $Pd_3X_4$ ($X$ = O, S) anionic frameworks that are ionically bonded to $A$ cations ($A$ = Na, La). These materials were recently identified computationally as harboring eightfold fermions. Here we report the preparation of single crystals and electronic properties of $LaPd_3S_4$. Measurements down to $T = 0.45$ K and in magnetic fields up to $\mu_0 H = 65$ T are consistent with normal Fermi liquid physics of a Dirac metal in the presence of dilute magnetic impurities. This interpretation is further confirmed by analysis of specific heat, magnetization measurements and comparison to density functional theory (DFT) calculations. Through a bonding analysis of the DFT electronic structure of $NaPd_3O_4$ and $LaPd_3S_4$, we identify the origin of the stability of the anionic $Pd_3X_4$ framework at higher electron counts for $X$ = S than $X$ = O, and propose chemical tuning strategies to enable shifting the 8-fold fermion points to the Fermi level.

## Introduction

The emergence of exotic electronic properties in extended solids is governed by the chemical bonding and symmetries that are present. Understanding how local chemical motifs give rise to interesting features in band structures is essential to the design of theoretically predicted, and often quantum, states of matter.[1-20] Elucidating how the atomic arrangement and electron count stabilize structures is essential in experimentally realizing such materials. Symmetry has long been a key tool to classify the mapping between local chemistry and band structure. Recently, the field of topology has added to these additional considerations to enable a full understanding of the resultant electronic behavior of a material. Aided by the computational accuracy in predicting the topological features such as the $\mathbb{Z}_2$ indices and in confirming key features in band structures such as linear dispersions in Dirac materials or band inversions in topological insulators, recent work has established a voluminous set of chemical principles to design new families of topological materials.[4,9,11]

One type of topological behavior is particularly interesting from a chemical perspective. Without inclusion of spin-orbit coupling, the highest known degeneracy by point group symmetries in a non-linear molecule/structure is six, i.e. the six degenerate $t_{2g}$ spin-orbitals in an octahedral crystal field system. Inclusion of spin-orbit coupling and an explicit treatment of the electron spin, utilizing double groups, drops the maximum degeneracy down to four. However, it was recently recognized in the context of topology that specific non-symmorphic symmetries (consisting of a point symmetry operation plus a fractional lattice translation), when paired with the appropriate non $k$=0 high symmetry point in the Brillouin zone could enforce an additional degeneracy not reflected in a double point group. Specifically, the cross product of two irreducible representations (irreps) or the cross product of one irrep with itself could force six or eight bands to intersect at a non-zero high symmetry point in the *k* space. In the case of cubic systems, at the high symmetry point, due to the non-symmorphic symmetry (NSS) in the space group, the 4-fold screw rotation and either the inversion center and translation or just the inversion center create a situation in which eight bands are degenerate.[8,10,21-23] These states, much like Dirac fermions, have the protection of both time-reversal and inversion symmetry, as shown in Table-1.[21-23]

Recent computational work has identified the palladium bronzes $NaPd_3O_4$ and $LaPd_3S_4$ as hosting such eight-fold fermion states.[8,11] Both crystallize in cubic symmetry space group *Pm-3n* that has the required symmetries to host these highly degenerate states at the **R**=(½, ½, ½) point of the Brillouin zone.[8,23] The structure consists of corner-sharing $PdX_4$ ($X$ = O,S) units that form three orthogonal but symmetry-related 1D chains of stacked $[PdX_4]$ units that together form an anionic framework that accepts charge from the $Na^+$/$La^{3+}$ cation. Both are known to be metallic, in agreement with computational predictions (and in contrast to other 8-fold fermion candidates such as $Bi_2CuO_4$).[8,10] A recent theoretical study elucidated in detail how topological states arise from the local chemical bonding in $NaPd_3O_4$.[11] Yet the existing literature is sparse on the crystal growth and low-temperature physical properties.[24] Here we report the single crystal growth, single-crystal structure, thermodynamic properties, and field-dependent transport properties of $LaPd_3S_4$. We find the behavior is that of a classic Dirac-like Fermi liquid modified by the presence of trace magnetic impurities, with no exotic signatures indicative of the presence of eight-fold fermions. We attribute this to the Fermi level not being suitably tuned to the 8-fold degenerate points. Using DFT calculations combined with crystal orbital Hamilton population (COHP) analysis, we identify the chemical bonding origins of the stability of the anionic $[Pd_3X_4]$ framework as a function of electron count and the identity of $X$, enabling us to propose future materials that may enable accessing novel electronic properties arising from 8-fold fermions.

## Methods

**Synthesis:** $LaPd_3S_4$ single crystals were grown using a long-term annealing synthesis. Stoichiometric amounts of La (Ames Laboratory 99.99%), Pd (BeanTown Chemical 99.95% trace metals basis), and S (Alfa Aesar, 99.999%, metals basis) with total mass of 0.5 grams were sealed in evacuated quartz ampoules. The ampoules were placed in a box furnace at an angle of 45° (with the sample charge close to the interiors of the furnace) and heated to $T$=900°C at rate=80°C/hr for 24 hours and subsequently heated to $T$=1100°C at a rate=80°C/hr and held for 28 days . The ampoules were quenched at $T$=1100°C and were allowed to cool in air. The ampoules were opened at room temperature and the obtained crystals were 1-3 mm in size. Phase purity and single-crystal domain size were characterized using powder, single crystal, and Laue x-ray diffraction.

**X-ray diffraction:** Powder x-ray diffraction (XRD) data, used to confirm phase purity of the single crystals, were collected over an angle range of 5°–60° on a laboratory Bruker D8 Focus diffractometer with a LynxEye detector and Cu Kα radiation. Single-crystal x-ray diffraction data were collected on a Bruker-Nonius X8 Proteum (Mo Kα radiation) diffractometer equipped with an Oxford cryostream. Integration and scaling were performed using CrysalisPro (Version 1.171.39.29c, Rigaku OD, 2017). A multiscan absorption correction was applied using SADABS.[31] The structure was solved by direct methods, and successive interpretations of difference Fourier maps were followed by least-squares refinement using SHELX and WinGX.[32,33] Laue diffraction was performed using a Multiwire Laboratories MWL110 diffractometer with a W source to check the orientation of the crystals and to identify possible twinning based on the spot sizes and shapes.

**Heat capacity:** The heat capacity data were collected on a Quantum Design Physical Properties Measurement System using the relaxation method (HC option) and a 1% temperature rise. The sample was mounted on the sample stage using Apiezon N grease.

**Magnetization:** The magnetization was carried out the in the Quantum Design Magnetic Properties Measurement System (MPMS3). Magnetization as a function of temperature at $\mu_0 H$ =1T at $T$=2-300K and magnetization as a function of magnetic field $\mu_0 H$ =-7 to 7 T were carried out respectively.

**Electrical transport measurements:** Resistivity measurements were carried out in a physical property measurement system (PPMS-12T, Quantum Design) using the Resistivity option. For longitudinal resistivity, ohmic contacts were made on the naturally grown crystals with silver paint and 25 μm platinum wires. The longitudinal and Hall resistivity were measured in 4-point Kevin probe technique, respectively using a current of 3.0–5.0 mA at a temperature range from 0.45 to 300 K and magnetic fields up to 12 T. The high pulsed field-dependent resistivity was measured in a four-point geometry using a 67 T non-destructive pulsed magnet driven by a capacitor bank at the Dresden High Magnetic Field Laboratory. The excitation current was 1 mA with a frequency of 3333 and 7407 Hz.

**Computational methods**: Density functional theory calculations were carried out using Quantum Espresso 7.0 with the PBE functional and PAW (Projector Augmented Wave) pseudopotentials available at http://pseudopotentials.quantum-espresso.org/legacy_tables.[34] Kinetic energy cutoffs of 75 Ryd and 300 Ryd were used for wavefunctions, and charge density/potential respectively. Gaussian smearing of 0.14 eV with an energy convergence threshold of $10^{-9}$ Ryd and a 15x15x15 *k*-point mesh were used during the SCF (Self-consistent field) process. LOBSTER 4.1.0 was used to carry out the orbital resolved density of states and COHP analysis. Rotation of orbitals into the local crystal field on a per-atom basis was used.

## Results

**Crystal and band structure of rare-earth palladium sulfides.** $LaPd_3S_4$ is reported to crystallize in a three-dimensional non-symmorphic $NaPt_3O_4$ structure in space group *Pm-3n* (No: 223). In this cubic structure, Figure-1(a,b), the Pd and *X* form square planar units with a three dimensional connectivity, and rare earth and *X* form body-centered polyhedra with the rare earth atom in the center and eight *X* atoms in the corner edges. The square planar units are aligned along the principle axes of the unit cell, and form 1D chains with a nearest neighbor Pd-Pd spacing of ~2.8 Å ($NaPd_3O_4$) and ~3.4 Å ($LaPd_3S_4$). This crystal structure is consistent with the morphology of the as-grown single crystals which are cuboctahedral and 3-dimensional. It is also consistent with Laue diffraction of a polished slice, Figure-1(c,d), which shows clean spots indexed as the [112] crystal orientation. Further, analysis of single crystal X-ray diffraction data collected at $T$=213 K, Table-2, 3, and 4, are in agreement with the previously reported structure for $LaPd_3S_4$.[35]

Figure-2(a,b) shows a comparison of the band structures of $NaPd_3O_4$ and $LaPd_3S_4$ obtained from non-magnetic DFT calculations with the PBE functional. These band structures are in agreement with prior reports, and consistent with chemical intuition that the alkali/rare earth cation are ionically bonded, resulting in band structures dominated by the $Pd_3X_4$ framework around $E_F$.[8] As per the cross product of irreducible representations $\Gamma_5(4)$, $\Gamma_6 \oplus \Gamma_7(8)$, multiple eight-fold crossings at the **R** point in the Brillouin zone are observed above and below $E_F$ in both materials.[8,23] The two band structures are qualitatively very similar, with a higher Fermi level in $LaPd_3S_4$ due to the increase in charge transfer to the anionic framework on changing from $Na^{1+}$ to $La^{3+}$. There are some notable differences, however. Across the Brillouin zone, some of the states derived from Pd *d* orbitals are shifted upwards, closer to the circled 8-fold fermion point, in $NaPd_3O_4$ relative to $LaPd_3S_4$. At the **R** point, the energy difference between the two 8-fold degenerate points

closest to $E_F$ is 1.5 eV in $NaPd_3O_4$ but increases to 1.9 eV in $LaPd_3S_4$. $E_F$ is approximately equidistant in energy from both in $NaPd_3O_4$, but in $LaPd_3S_4$ $E_F$ is only ~0.1 eV below the upper 8-fold degenerate point. Along the R-X direction, another substantial difference occurs for the lowest state coming from the 8-fold fermion just above $E_F$: in $NaPd_3O_4$, the band monotonically decreases in energy with a partial saddle point in the middle, whereas in $LaPd_3S_4$ the band first decreases in energy before ultimately rising above the 8-fold point before reaching X.

Given these similarities despite the large difference in electron count, it is natural to ask what allows for this wide stability range. Inspection of known materials, shown in Table-3, shows that indeed the anionic frameworks form with corner-sharing cations ranging from +1 to +3. The known compounds are not evenly distributed between $X$ = O and $X$ = S. Instead, it appears that when $X$ = O, cations transferring up to two electrons per $Pd_3O_4$ unit are stable (including the 0-electron "cation vacancy" $Pd_3O_4$ binary, implying stable average Pd valences of 2.66+ to 2+.[11] On the other hand, when $X$ = S, cations must transfer more than two electrons per $Pd_3S_4$ unit for a stable structure to result, i.e. average Pd valences < 2+, Table-5. Does this trend reflect an underlying principle of bonding, or is it simply the consequence of a bias in the set of known materials?

To answer this question, we carried out an orbital-resolved density of states (DOS) and COHP analysis of the $NaPd_3O_4$ and $LaPd_3S_4$ band structures. The orbitally projected DOS, shown in Figure-3(b,d), show some important differences. Both have bands of $d$-orbitals consistent with expectations from a crystal field theory treatment of square planar $PdX_4$ units combined with the connectivity – a very diffuse $d_{x^2-y^2}$band and a less dispersive $d_{xy}$band. The $d_{xz}/d_{yz}$ and $d_{z^2}$ bands are significantly different, however: in the case of $LaPd_3S_4$, they appear to be positioned roughly as expected, with narrow bandwidths indicating little overlap between adjacent [$PdS_4$] units. In contrast, in $NaPd_3O_4$, these bands are more dispersive, and shifted in energy as expected if there was strong Pd-Pd bonding between square planes along the 1D chains. These differences arise from the change in Pd-Pd distance from ~2.8 Å in $NaPd_3O_4$ (close to the Pd-Pd distance of 2.75 Å in Pd metal), to ~3.4 Å in $LaPd_3S_4$. This has a significant consequence on the energetics: starting from $[Pd_3X_4]^{2-}$, the energy costs, as estimated from the ICOHP in Figure-2(e,f), of adding electrons is more than twice as large for $X$ = O than $X$ = S. This is because the stronger intrachain interactions make the orbitals just above or below the 2+ balance point more antibonding in character for $X$ = O. This nicely explains why electron counts lower than 2+ are more prevalent for $X$ = S than for $X$ = O. Why formal oxidation states of 2 and greater are not found in the case of $X$ = S is not apparent from this analysis, but is likely a consequence of the stability of polyanionic sulfide units under such oxidizing scenarios (these typically do not form with oxygen).

This analysis suggests that reaching the 8-fold fermion point above $E_F$ in $NaPd_3O_4$ will be extremely challenging chemically, as it requires additional electron transfer that further destabilizes the structure. Since the 8-fold point is very close in $LaPd_3S_4$, it may be that electrostatic gating or applying pressure or partial substitution of $La^{3+}$ by a tetravalent cation (e.g., $Th^{4+}$), will be successful in raising the chemical potential to the 8-fold crossing.

Another intriguing possibility suggested by these results is the possibility to engineer both $e^-$ and $h^+$ carrying Pd catalysts, since the $[Pd_3X_4]^{2-}$ charge state is the point of cross-over from hole-like to electron-like behavior. Various forms of $APd_3O_4$ (and the Pt versions) have been extensively explored for potential use in catalysis/electrocatalysis, and this is one structure thought to be important in Adam's catalyst (a hydrogenation catalyst often represented as hydrated platinum dioxide, but in reality a substantially more complicated with various platinum bronzes forming in situ during catalyst use).[11] Given the recent forays into the use of topological states of matter to improve catalysis, an interesting line of future work would be to explore whether the additional electron count tunability enabled by the replacement of O by S enables additional catalytic behavior or performance.

**Heat capacity of $LaPd_3S_4$, a Pauli paramagnet** As expected for a Pauli paramagnet, $LaPd_3S_4$ does not undergo a phase transition associated with magnetic order. The resulting heat capacity data shown in Fig. 4a shows a plot of $C_p/T$ vs $T^2$, with a linear fit $T$=2-7 K . The linear fit follows the equation $C_p/T = \beta_3 T^2 + \gamma$. Where $\beta_3$ is the phonon contribution and $\gamma$ is the Sommerfield coefficient associated with the electronic contribution in the heat capacity. The extrapolated values from the linear fit are $\beta_3$=0.900 mJ.mol$^{-1}$.K$^{-4}$ and $\gamma$=20.5 mJ.mol$^{-1}$.K$^{-2}$ respectively. Sommerfield coefficient corresponds to a density of states at $E_F$ of 9 states/eV/cell. This is approximately twice as large as the value obtained from DFT (5 states/eV cell). This most likely arises due to a combination of electron-phonon coupling, spin-spin correlations (as in Pd metal) and mass enhancement due to interaction with magnetic impurities (vida infra). The $\beta_3$ parameter corresponds to a Debye temperature of $T_D$ = 87 K. This is smaller than expected given the atomic weights of the elements involved. The $\beta_3$ parameter corresponds to a Debye temperature of $T_D$ = 87 K. This is smaller than expected given the atomic weights of the elements involved. To explore this in greater detail, we carried out heat capacity measurements to higher temperature, $T$ = 270 K, as shown in Fig. 3b. These data show $LaPd_3S_4$ reaches the Dulong-Petit limit around $T$ = 200 K. A detailed analysis of the phonon contributions (see SI) does not reveal any unexpected behavior.

**Transport behavior in $LaPd_3S_4$** Figure-4 summarizes the transport measurements and analysis on $LaPd_3S_4$. Figure-4(a) shows the temperature-dependent resistivity of a "thick" specimen. The behavior is indicative of a metal, with a residual resistivity ratio (*RRR*) *RRR* = 17, and a monotonically decreasing resistivity. The residual resistivity is 1 mΩ-cm, larger than in many metals, but not as high in comparison to the six-fold double Dirac material $WP_2$. The relatively high value is suggestive of either a low carrier concentration or low mobility. As prior Hall measurements indicate a high number of carriers, ~1.7 electron carriers per unit cell, or $5.6\cdot10^{21}$ carriers/cm$^3$ (consistent with the DFT DOS prediction) this implies low carrier mobility (for a metal) of ~100 cm$^2$/V/s at $T$=2 K.[24] This is consistent with the presence of significant impurity scattering (*vida infra*). One unusual feature is the apparent $T$-linear resistivity. Often attributed to "strange metal" behavior, such linear dependence now known to arise when linearly dispersing (Dirac) states are present, such as in Kagomè metals and Dirac semimetals.[8,22] Given linear crossings at $E_F$ are observed in the DFT calculations; we attribute this T-linear behavior to the Dirac states predicted to exist in $LaPd_3S_4$.

Given the non-magnetic composition, as expected the magnetoresistance (MR) does not show any sign of magnetically induced transitions up to $\mu_0 H$ = 67T as shown in Figure-4(b). However, below T ~ 2 K, there is a slight negative MR, Figure-4(c). The origins of this anisotropic MR could be due to shape effects in the Fermi surface of $LaPd_3S_4$, but is most likely due to presence of magnetic impurities (*vida infra*). The presence of such defect scattering is also suggested by the lack of observable quantum oscillations.

Some quantitative information can be extracted from the MR data. Using a general model for a two channel metal, the MR is expected to behave as $\rho/\rho_0 = 1 + [(1 + bB^2)/(1 + cb^2B^2)]$, where $\mu_0 H$ is the applied magnetic field, and b and c are coefficients depending on the ratio of carrier concentrations and carrier mobilities of the two channels.[36] These constants can in turn be related to estimated carrier densities and mobilities as a function of temperature as shown in Figure-5(d,e). The temperature trends are as expected for a normal metal, with no evidence for exotic physics arising from 8-fold fermions.

All of the above analysis suggests the presence of significant impurity scattering in $LaPd_3S_4$ single crystals. Magnetization measurements can be used to quantify the number of magnetic scattering centers. Magnetic susceptibility, estimated as *M*/*H*, is shown in Figure-6(a). The behavior is that of a Pauli paramagnet, with a definite Curie-type upturn at low temperatures as shown in SI Figure-1. Initial attempts to fit to the full Curie-Weiss law gave Weiss temperature magnitudes of 70+ K, unphysical given the nonmagnetic nature of the constituent elements. Thus we instead carried out the analysis assuming that the Weiss temperature is equal to zero. This analysis, Figure-6(a) yields a Curie constant of $C$ = 0.0111 emu.K.(mol.La)$^{-1}$.Oe$^{-1}$, and a temperature-independent susceptibility of $\chi_0 = 1.14\cdot10^{-4}$ emu.(mol.La)$^{-1}$.Oe$^{-1}$. Taking into account core diamagnetism, the temperature independent susceptibility is $3.1\cdot10^{-4}$ emu.(mol.La)$^{-1}$.Oe$^{-1}$, about twice that predicted for a simple metal from the density of states: $\chi_{calc} = \mu_B^2\, g(\varepsilon_F) = 1.5 \text{ x } 10^{-4}$ emu mol$^{-1}$ Oe$^{-1}$. This is in agreement with a similar enhancement of the specific heat linear value. The Curie constant, reflective of the unpaired electrons, would correspond to 0.0111/0.375 ~ 3% of $S$=1/2 impurity spins, a high number given the purity of the starting reagents. If the dominant magnetic impurity were instead $Ni^{2+}$ (a common contaminant of Pd), then we would expect $S$=1 and this would correspond to 0.0111/1 ~ 1.1% of Ni for Pd. On the other hand, if the dominant magnetic impurity is magnetic rare earths in place of La, then the estimated fraction (for $Nd^{3+}/Pr^{3+}$, common impurities in La) is 0.0111/1.62 ~ 0.69%. The low temperature *M*(*H*) data, shown in Figure-6(b), let us constrain these possibilities. The saturation magnetization is 0.017 $\mu_B$/f.u. For spin-1/2 impurities, this corresponds to 0.017/1 = 1.7% $S$=1/2, not a good match to the Curie analysis. For $Ni^{2+}$ impurities, this corresponds to 0.017/2 ~ 0.9% of Ni, within 25% of that estimated from the Curie constant. On the other hand, for $Nd^{3+}/Pr^{3+}$, this corresponds to 0.017/3.13 ~ 0.54%, also within 25% of that from the temperature-dependent analysis. We thus ascribe the dominant magnetic impurity to either Ni for Pd, magnetic rare earths substituted for La, or both.

The 0.5% magnetic rare earth impurities is consistent with both the magnitude of the residual resistivity and the enhanced low temperature specific heat $\gamma$ term. For dilute, magnetic Mn doped into simple metals such as Ag, Au, and Cu, 0.5% of high spin impurities reduce carrier mobility by a factor of ~10. Thus in the absence of these magnetic impurities, we would expect the residual resistivity to be a factor of 10 lower, increasing *RRR* above 100 and yielding an estimated carrier mobility of ~1000 cm$^2$/V/s, both reasonable values for high but not extreme purity metals.[37] The magnitude of the *T*-linear specific heat is more difficult to estimate, but if we use dilute Mn in Ag as a reference, the low-temperature *T*-linear contribution from the impurities is expected to be ~1 J.(mol-impurity)$^{-1}$.K$^{-2}$. Thus, we would estimate ~5 mJ.mol-f.u.K$^{-2}$ to arise from the magnetic impurities.[38] The fitted $\gamma$=20.5 mJ.mol$^{-1}$.K$^{-2}$ from heat capacity in Figure-3(a) are very close to what is expected: $\gamma_{cal} = \frac{g(\varepsilon_F)}{424.25} + impurities$ = 11.8 mJ.mol$^{-1}$.K$^{-2}$ + 5 mJ.mol$^{-1}$.K$^{-2}$ = 16.8 mJ.mol$^{-1}$.K$^{-2}$. We cannot rule out additional contributions from the presence of high concentrations of sulfur point defects in addition to the magnetic dopants from Pd and La.

## Conclusion

$LaPd_3S_4$ single crystals prepared via long term annealing and grain growth technique crystallize in the cubic *Pm-3n* $NaPt_3O_4$ structure type and exhibit metallic behavior. The heat capacity results show no phase transitions as expected with $C_p/T$ vs $T^2$. At low temperature, the $C_p/T$ vs $T^2$ is modeled to a linear fit with extrapolated electronic and phononic contributions $\gamma$=20.5 mJ.mol$^{-1}$.K$^{-2}$ and $\beta_3$=0.900 mJ.mol$^{-1}$.K$^{-4}$. The $\gamma$ value is in the range of what one might expect for a metal in the presence of dilute magnetic impurities or with high concentration of point defects such as sulfur. The DFT and transport results confirm the metallic nature. As predicted due to symmetry operations, $LaPd_3S_4$ has a double Dirac crossing at the R point of the Brillouin zone, these states are primarily dominated by the square planar Pd-S framework which is also seen in the $NaPt_3O_4$ structure. The transport properties up to 65T show the metallic and non-saturating MR without any appreciable perturbation due to magnetic oscillations. All of the data is consistent with the presence of ~0.5% magnetic rare earth impurities on the La and/or Pd site. Our bonding analysis comparing $NaPd_3O_4$ to $LaPd_3S_4$ identifies the chemical origin of the stability of high Pd valences in the case of the oxide, and low Pd valences in the case of the sulfide. Together, these results demonstrate that if 8-fold fermions generate exotic physics, additional chemical tuning and purification is required to observe it.

**Acknowledgments:** The authors would like to thank M. Siegler and R. Koban for technical assistance. This work was supported as part of the Institute for Quantum Matter and Energy Frontier Research Center, funded by the U.S. Department of Energy, Office of Science, Office of Basic Energy Sciences, under Award DE-SC0019331. This work was also financially supported by the European Research Council (ERC Advanced Grant No. 742068 'TOPMAT'). We also acknowledge funding by the DFG through SFB 1143 (project ID. 247310070) and the Würzburg-Dresden Cluster of Excellence on Complexity and Topology in Quantum Matter ct.qmat (EXC2147, project ID. 39085490). Quantum Espresso. Place where DFT was run on. We acknowledge the support of the HLD at HZDR, a member of the European Magnetic Field Laboratory (EMFL).

| Fermions | Idea | Time Reversal Symmetry | Inversion symmetry | Schematic | Example |
|---|---|---|---|---|---|
| Dirac<br>(4 fold) | "massless" spin-degenerate electrons<br>$E \propto k$ | Preserved | Preserved | 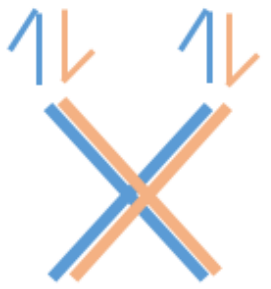<br>Doubly degenerate | $Cd_3As_2$<br>graphene<br>$Na_3Bi$ |
| Weyl<br>(2 fold) | "massless"<br>non-spin-degenerate<br>(1 Dirac = 2 Weyl's of opposite chiralities. The 2 Weyl's are degenerate) fermions | Type 1 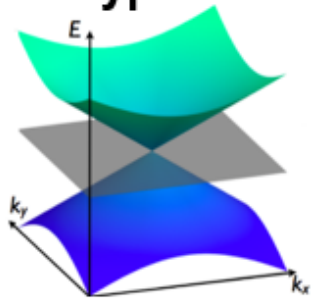 Type 2 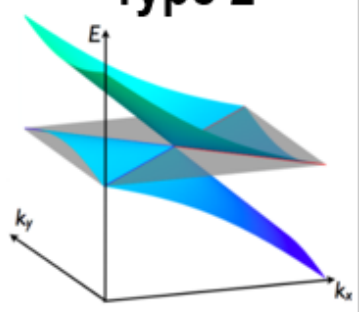<br>Breaks either inversion or TR symmetries | | 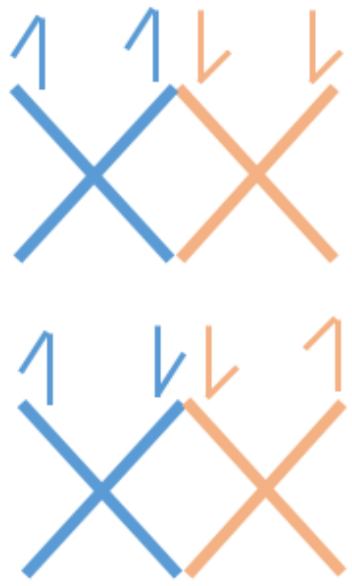 | TaAs<br>NbAs |
| Double<br>Dirac<br>(8 fold) | Non symmorphic symmetry is essential for stabilizing eight-fold degenerate points | Preserved | Preserved | 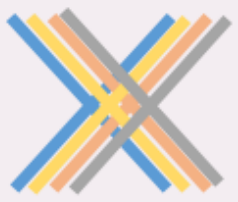 | $LaPd_3S_4$<br>$NaPd_3O_4$<br>$Ta_3Sb$ |

**Table-1** A table explaining Dirac, Weyl, and double Dirac fermions, the conceptual idea, time reversal symmetry dependence, inversion symmetry dependence, a schematic, and some examples.[11,25-30]

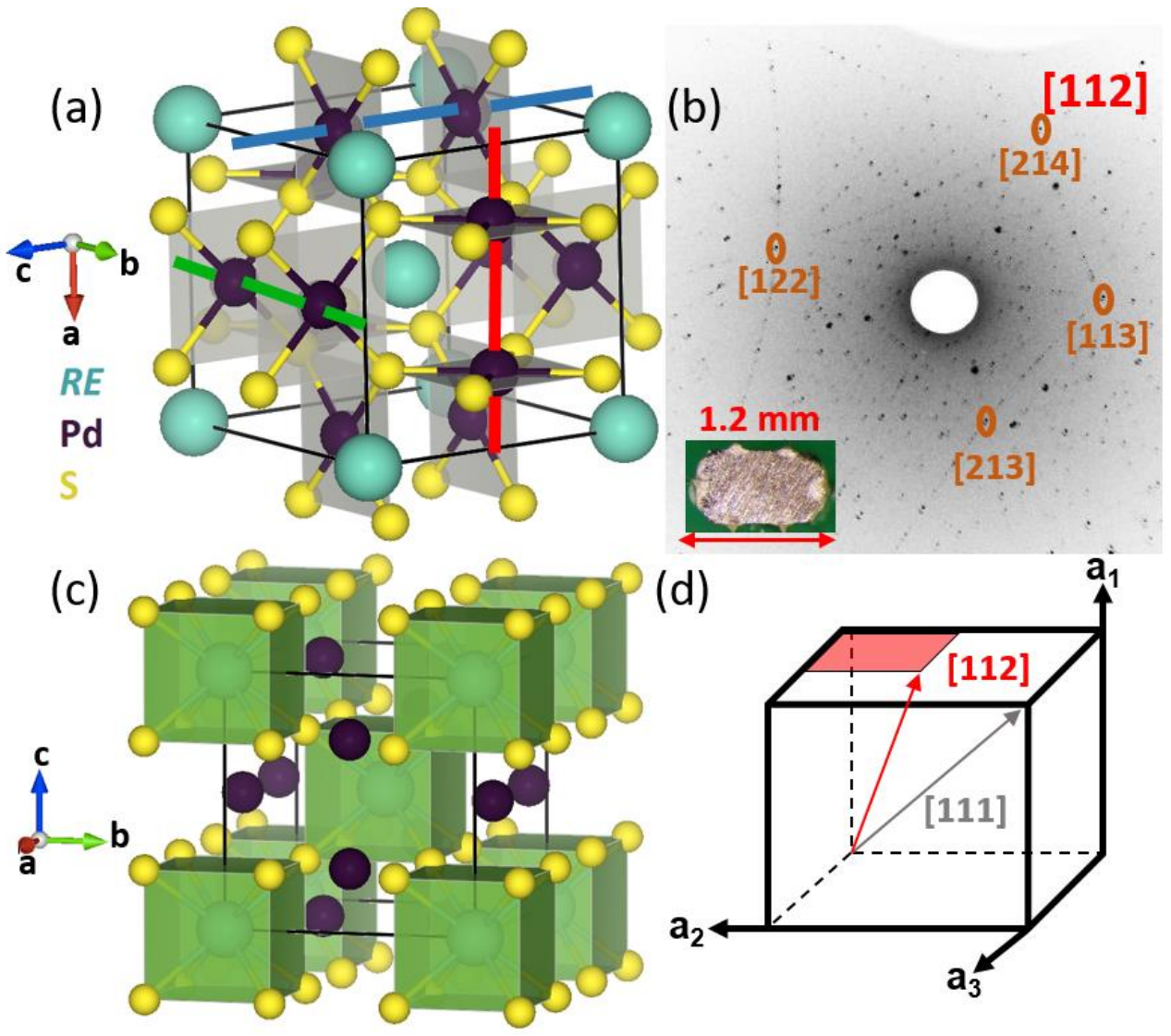

**Figure-1 (a)** Crystal structure of $REPd_3S_4$ crystalizing in *Pm3-n*. The figure shows square planar $[Pd_3S_4]^{3-}$ units. The red, blue, and green lines show Pd-Pd bonds that are connected in three dimensions with an average bond distance of ~2.8 Å, which depicts the pseudo-octahedral coordination of Pd. **(b)** A Laue image of the single crystal of $LaPd_3S_4$ along the [112] direction. The inset shows the single crystal of $LaPd_3S_4$**.** **(c)** Crystal structure of $REPd_3S_4$ crystalizing in *Pm-3n*. The figure shows square prismatic of RE and S units. **(d)** A cubic unit cell with depiction of the [112] miller plane. It is important to observe that the square plane and square prisms in **(a)** and **(c)** are eclipsed to one another instead of staggered.

| **Formula** | **$LaPd_3S_4$** |
|---|---|
| Crystal system | Cubic |
| Space Group | *Pm*-*3n* (No. 223) |
| *a* (Å) | 6.73384(7) |
| V (Å$^3$) | 305.343(49) |
| Z | 2 |
| M/g mol$^{-1}$ | 586.366 |
| $\rho_0$/g cm$^{-3}$ | 6.370 |
| $\mu$/mm$^{-1}$ | 16.77 |
| Radiation | Mo K$\alpha$, $\lambda$= 0.71073 Å |
| Temperature (K) | 213 K |
| Reflections collected/number of parameters | 5116/86 |
| Goodness-of-fit | 1.604 |
| R[F][a] | 0.0111 |
| $R_w(F_o^2)$[b] | 0.00252 |

[a] $R(F) = \Sigma||F_o| - |F_c||/\Sigma|F_o|$

[b] $R_w(F_o^2) = [\Sigma w(F_o^2 - F_c^2)^2/\Sigma w(F_o^2)^2]^{1/2}$

**Table-2**. Single crystal x-ray diffraction (SXRD) parameters and refinement statistics. Please note, $\rho_0$ is density.

|  | Occ. | Wyckoff Positions | *x* | *Y* | *z* | SOF |
|---|---|---|---|---|---|---|
| La | 1 | 2a | 0.0000 | 0.0000 | 0.0000 | 0.04167 |
| Pd | 1 | 6c | 0.2500 | 0.5000 | 0.0000 | 0.12500 |
| S | 1 | 8e | 0.2500 | 0.2500 | 0.2500 | 0.16667 |

**Table-3**. Atomic coordinates for $LaPd_3S_4$ determined by SXRD.

|  | $U_{(1,1)}$ | $U_{(2,2)}$ | $U_{(3,3)}$ | $U_{(2,3)}$ | $U_{(1,3)}$ | $U_{(1,2)}$ | $U_{eq}$ |
|---|---|---|---|---|---|---|---|
| La | 0.00045 | 0.00045 | 0.00045 | 0.00000 | 0.00000 | 0.00000 | 0.00045 |
| Pd | 0.00688 | 0.00066 | 0.00066 | 0.00000 | 0.00000 | 0.00000 | 0.00273 |
| S | 0.00289 | 0.00289 | 0.00289 | -0.00015 | -0.00015 | -0.00015 | 0.00289 |

**Table-4**. Anisotropic displacement parameters for $LaPd_3S_4$ determined by SXRD.

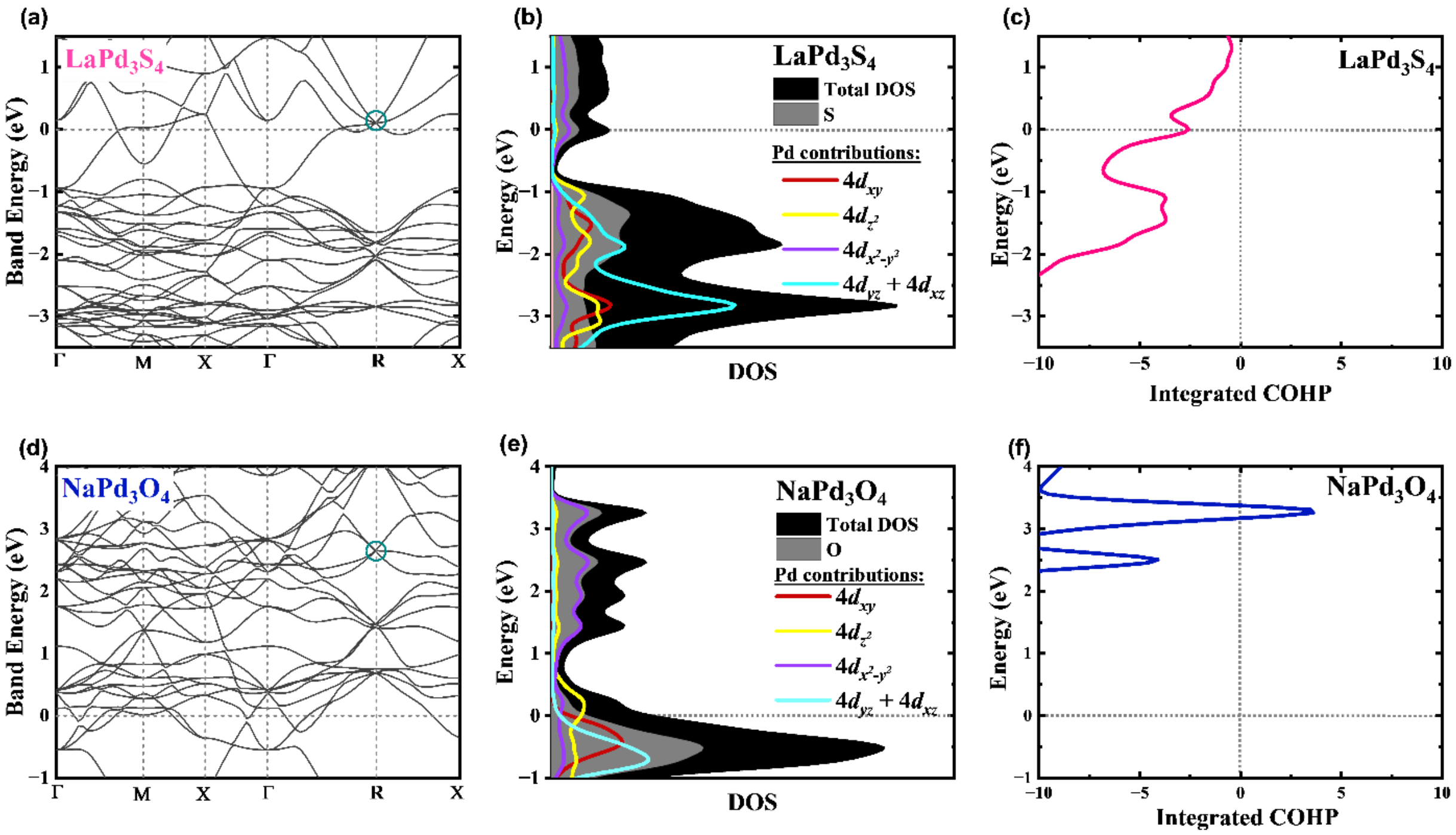

**Figure-2 (a)** Band structure of $LaPd_3S_4$ without spin orbit coupling (SOC), **(b)** density of states (DOS) with individual orbital contribution of Pd and S units, and **(c)** Integrated COHP analysis on $LaPd_3S_4$. **(d)**, **(e)**, and **(f)** for $NaPd_3O_4$ with the same contributions as $LaPd_3S_4$.

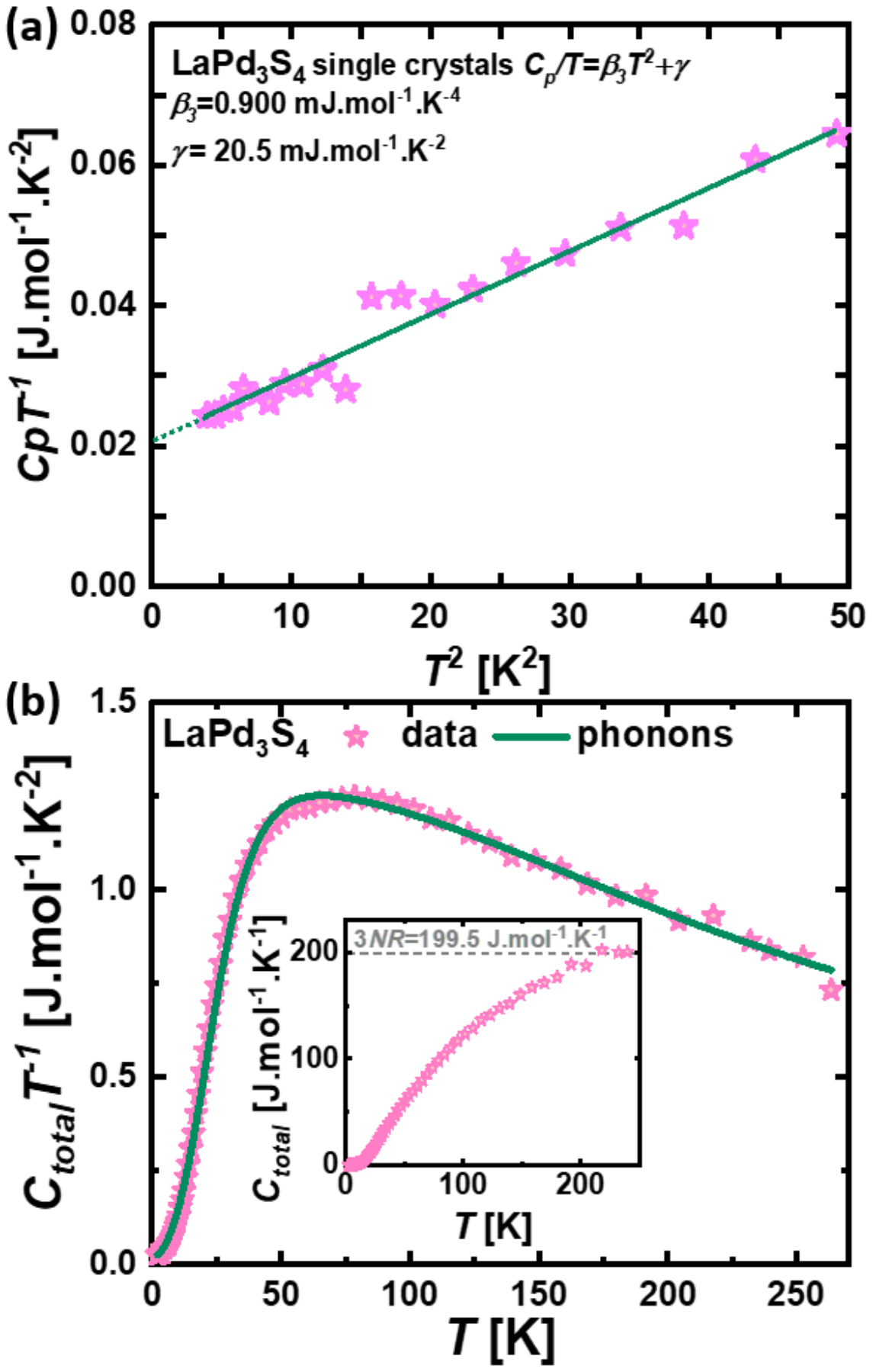

**Figure-3 (a)** Heat capacity divided by temperature as a function of temperature for $LaPd_3S_4$ at $\mu_o H$=0.1 T from $T^2$ ≈4-50 K. The purple asterisks are the data points, and the green line is the linear fit. The extrapolated values are the $\gamma$=20.5 mJ.mol$^{-2}$.K$^{-2}$ and $\beta_3$=0.900 mJ.mol$^{-1}$.K$^{-4}$. **(b)** The purple flowers are the data points, and the green is the phonons modeled using the 2-Debye term models from $T$≈2–270 K. The inset shows that the Dulong-Petit theoretical value of 3$NR$=199.5 J.mol$^{-1}$K$^{-1}$ is reached around $T$≈270 K.

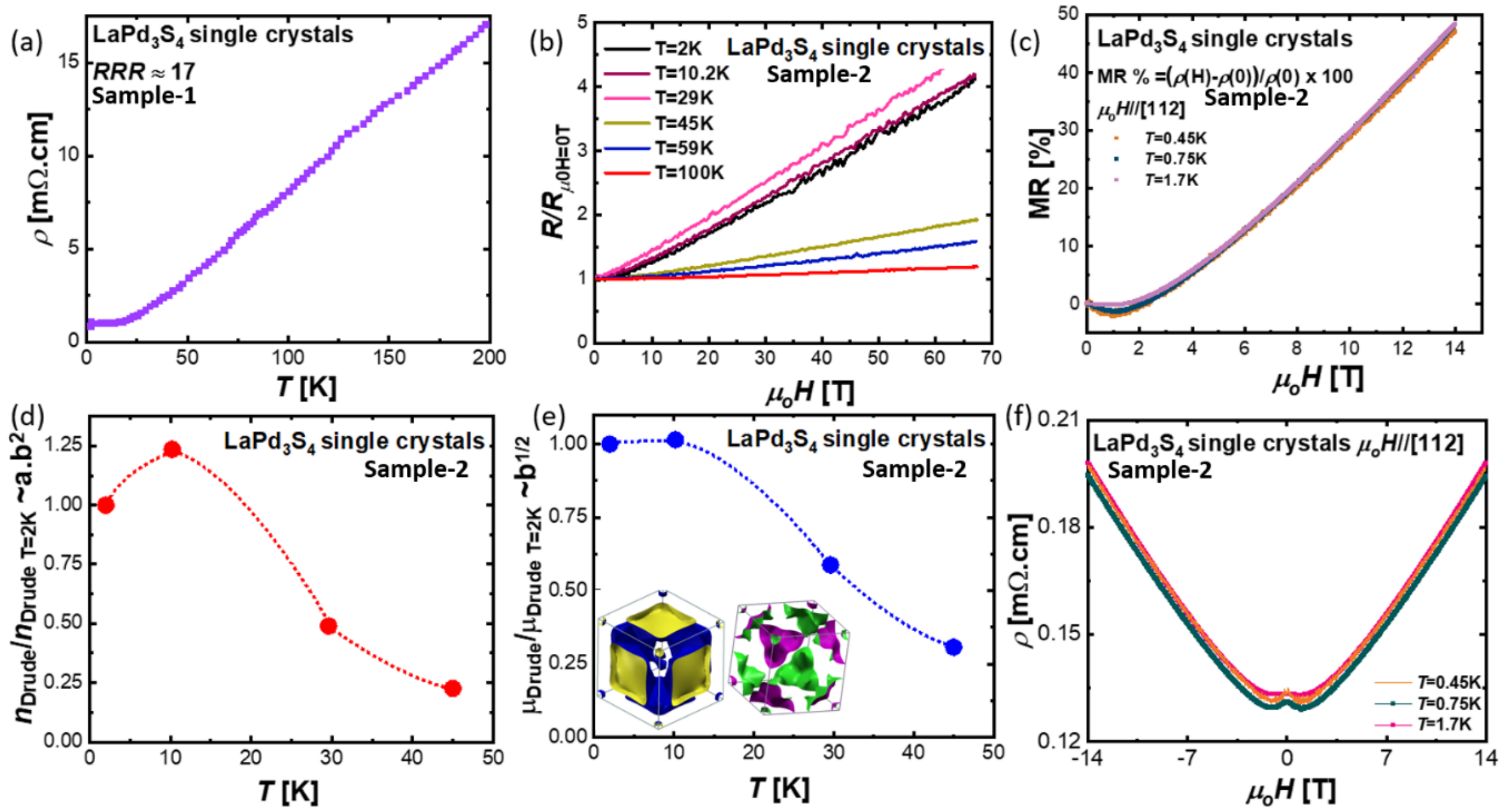

**Figure-4** **(a)** Resistivity of $LaPd_3S_4$ single crystals at $\mu_0 H$=0.1 T from $T$≈2–200 K. The residual resistivity ratio (*RRR*), *RRR*≈17. **(b)** Normalized resistance of $LaPd_3S_4$ single crystals as a function of magnetic field, where $\mu_0 H$≈0.1–67 T at $T$=2, 10.2, 29, 45, 59, and 100 K respectively. The trend is linear without any signs of magnetic oscillations. **(c)** The MR of $LaPd_3S_4$ single crystals with *RRR*≈50. There is a slight negative MR attributed to the small magnetic impurities in place of La. **(d)** The normalized carrier concentration of $LaPd_3S_4$ single crystals **(e)** The normalized carrier mobility of $LaPd_3S_4$ single crystals, and **(f)** The resistivity as function of magnetic field of $LaPd_3S_4$ single crystals at T=0.45, 0.75, and 1.7K respectively. Please note that the inset in **(e)** represents the Fermi surfaces close to $E_F$ in $LaPd_3S_4$ computed from DFT. These Fermi surfaces are large, which explains the lack of oscillations in the transport results. Also, the Sample-1 and Sample-2 are from the same batch crystal. Sample-2 is cut from a larger crystal of Sample-1.

| *Pm*-3*n* *M*$Pd_3$*X*$_4$ | Corner sharing cation | Pd oxidation state (nominal) | Pd-X (Å) | Lattice Parameter (Å) | References |
|---|---|---|---|---|---|
| $NaPd_3O_4$ | $Na^{1+}$ | 2.33+ | 1.9975 | 5.6479(6) | 11 |
| $CaPd_3O_4$ | $Ca^{2+}$ | 2+ | 2.032 | 5.747 | 39 |
| $SrPd_3O_4$ | $Sr^{2+}$ | 2+ | 2.060 | 5.826 | 39 |
| $LaPd_3S_4$ | $La^{3+}$ | 1.67+ | 2.383 | 6.7394(1) | 35, this work |
| $NdPd_3S_4$ | $Nd^{3+}$ | 1.67+ | 2.364 | 6.6864(4) | 24 |
| $EuPd_3S_4$ | ½ $Eu^{2+}$ and ½ $Eu^{3+}$ | 1.83+ | 2.3605 | 6.6765(5) | 24 |
| $CePd_3S_4$ | $Ce^{3+}$ | 1.67+ | 2.373 | 6.7130(1) | 40 |
| $ErPd_3S_4$ | $Er^{3+}$ | 1.67+ | 2.342 | 6.6240(1) | 35 |
| $HoPd_3S_4$ | $Ho^{3+}$ | 1.67+ | 2.346 | 6.6346(1) | 35 |
| $PrPd_3S_4$ | $Pr^{3+}$ | 1.67+ | 2.368 | 6.6989(1) | 35, 41 |
| $SmPd_3S_4$ | $Sm^{3+}$ | 1.67+ | 2.358 | 6.6692(1) | 35 |
| $YPd_3S_4$ | (1-x)$Y^{3+}$ and x$Y^{2+}$ [where x is an ambiguous number] | 1.67+ and 1.83+ | 2.346 | 6.6349(1) | 35, 42 |

**Table-5.** Key charge and structural parameter correlations in the *M*$Pd_3$*X*$_4$ structure type that crystalizes in *Pm*-3*n* (space group- 223 and where *M*= Na, Ca, Sr, La, Nd, Eu, Ce, Er, Ho, Pr, Sm, and Y and *X*= O and S).

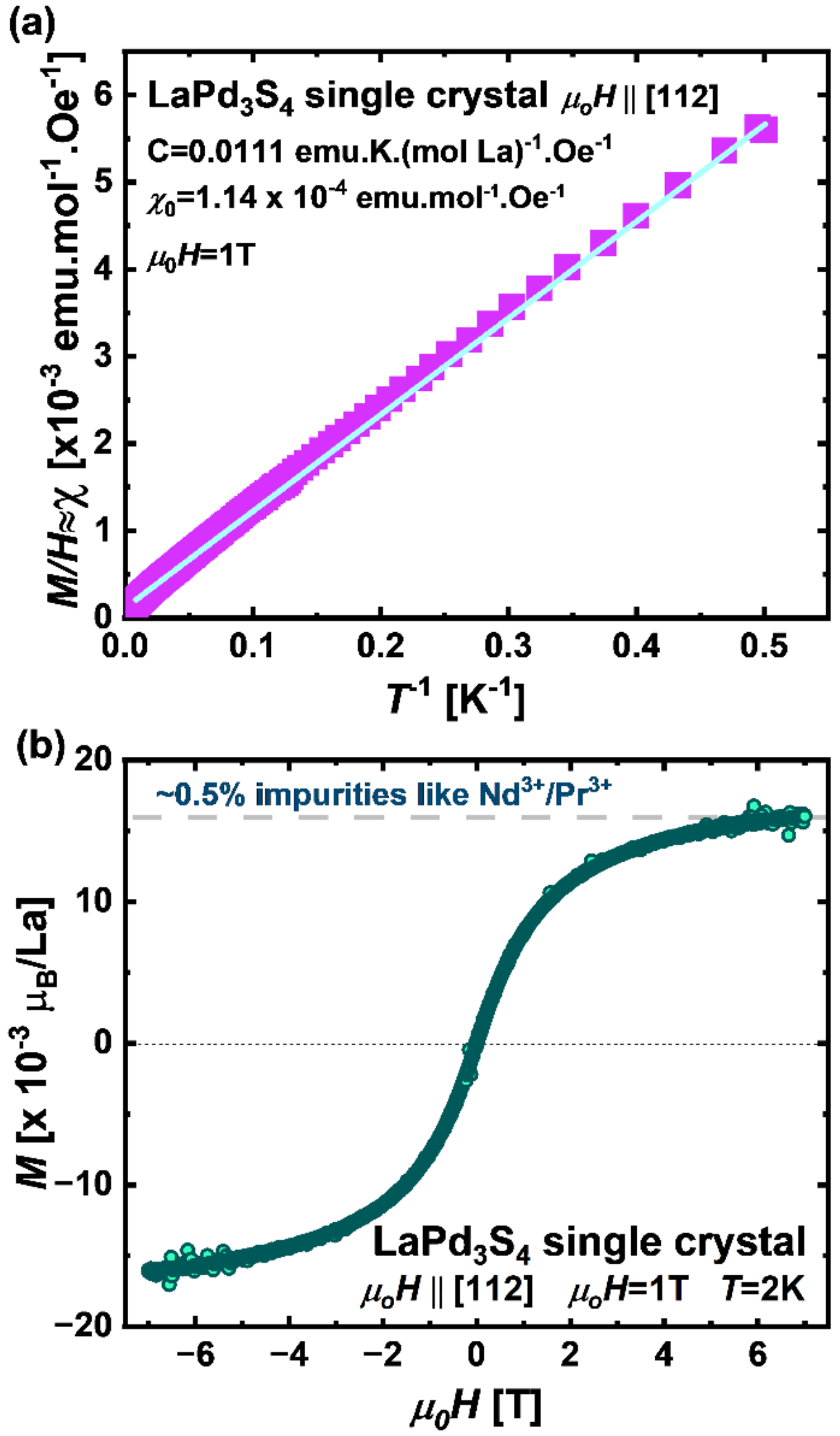

**Figure-5** **(a)** Curie analysis of the magnetization data. The magnetization is plotted as a function of $T^{-1}$ in $LaPd_3S_4$ single crystals at $\mu_o H$=1 T from $T$≈2–300 K. **(b)** Magnetization as a function of magnetic field temperature in $LaPd_3S_4$ single crystals at $T$≈2K and from $\mu_o H$=-7 to 7 T.

# Bonding and Electronic Nature of the Anionic Framework in $LaPd_3S_4$

Tanya Berry,[1,2*] Michael Nicklas,[3] Qun Yang,[3] Walter Schnelle,[3] Rafał Wawrzyńczak,[3] Tobias Förster,[4] Johannes Gooth,[3] Claudia Felser,[3] Tyrel M. McQueen[1,2,5*]

[1]*Department of Chemistry, The Johns Hopkins University, Baltimore, Maryland 21218, USA*
[2]*Institute for Quantum Matter, William H. Miller III Department of Physics and Astronomy, The Johns Hopkins University, Baltimore, Maryland 21218, USA*
[3]*Max Planck Institute for Chemical Physics of Solids, Nöthnitzer Straße 40, 01187 Dresden, Germany*
[4]*Hochfeld-Magnetlabor Dresden (HLD-EMFL) and Würzburg-Dresden Cluster of Excellence ct.qmat, Helmholtz-Zentrum Dresden-Rossendorf, 01328 Dresden, Germany*
[5]*Department of Materials Science and Engineering, The Johns Hopkins University, Baltimore, Maryland 21218, USA*

*Email: tberry@ucdavis.edu and mcqueen@jhu.edu

## Supporting Information Available:

Detail of the phonon heat capacity fitting parameters and magnetization results are described here.

**Phonon model to heat capacity of $LaPd_3S_4$**

To elucidate the phonon contributions further, the heat capacity data from $T$=2-270 K in $LaPd_3S_4$ single crystals was collected. Debye modes represent highly dispersing phonon branches, and Einstein modes localized optic modes. The presence of localized Einstein contributions was ruled out due to the absence of a peak in *a* $C_p/T^3$ vs $T$ plot.[15] Use of a single Debye contribution did not satisfactorily model the data. Thus, the heat capacity data was modeled with two Debye modes as well as an electronic contribution, with final parameters given in Table-SI1. The absence of an Einstein mode and lack of signatures of a proper phase transition suggest that the cubic structure of $LaPd_3S_4$ is retained at all temperatures up to room temperature, and that there are no breathing modes in the Pd-S and La-S coordination, or any other type of charge order associated with the exotic valence of Pd in the $NaPt_3O_4$ structure type.

$$\frac{C_p}{T} = \frac{C_D(\theta_{D1}, s_1, T)}{T} + \frac{C_D(\theta_{D2}, s_2, T)}{T} \qquad (1)$$

$$C_D(\theta_D, T) = 9sR\left(\frac{T}{\theta_D}\right)^3 \int_0^{\theta_D/T} \frac{(\theta/T)^4 e^{\theta/T}}{[e^{\theta/T}-1]^2} d\frac{\theta}{T} \qquad (2)$$

Where $\theta_D$ is the Debye temperature, $s_D$ is the oscillator strength, R is the molar Boltzmann constant, and $\gamma$ is the electronic contribution. The heat capacity also reaches the Dulong Petit limit at $T$=270K which below the glass freezing point of Apiezon N grease as seen in Fig. 3b inset.[1] The resulting model is in physical agreement with the total number of Debye oscillators adds up to 9.3(4), which is close to the total number of atoms per formula unit. The slightly large oscillator weight is due to large error bars of the fit parameters which is expected from the precision of the experimental data.

| $s_{D1}$ (oscillator strength/formula unit) | $s_{D2}$ (oscillator strength/formula unit) | $\theta_{D1}$(K) | $\theta_{D2}$(K) | $\gamma$ (mJ.mol$^{-1}$.K$^{-2}$) |
|---|---|---|---|---|
| 3.9(2) | 5.4(2) | 186(3) | 613(36) | 20.5(2) |

**Table-SI1.** Fitting parameters to the $C_p/T$ as a function of $T$ for $LaPd_3S_4$ to extract the phonon contribution. The data was fit from $T$=2-270K.

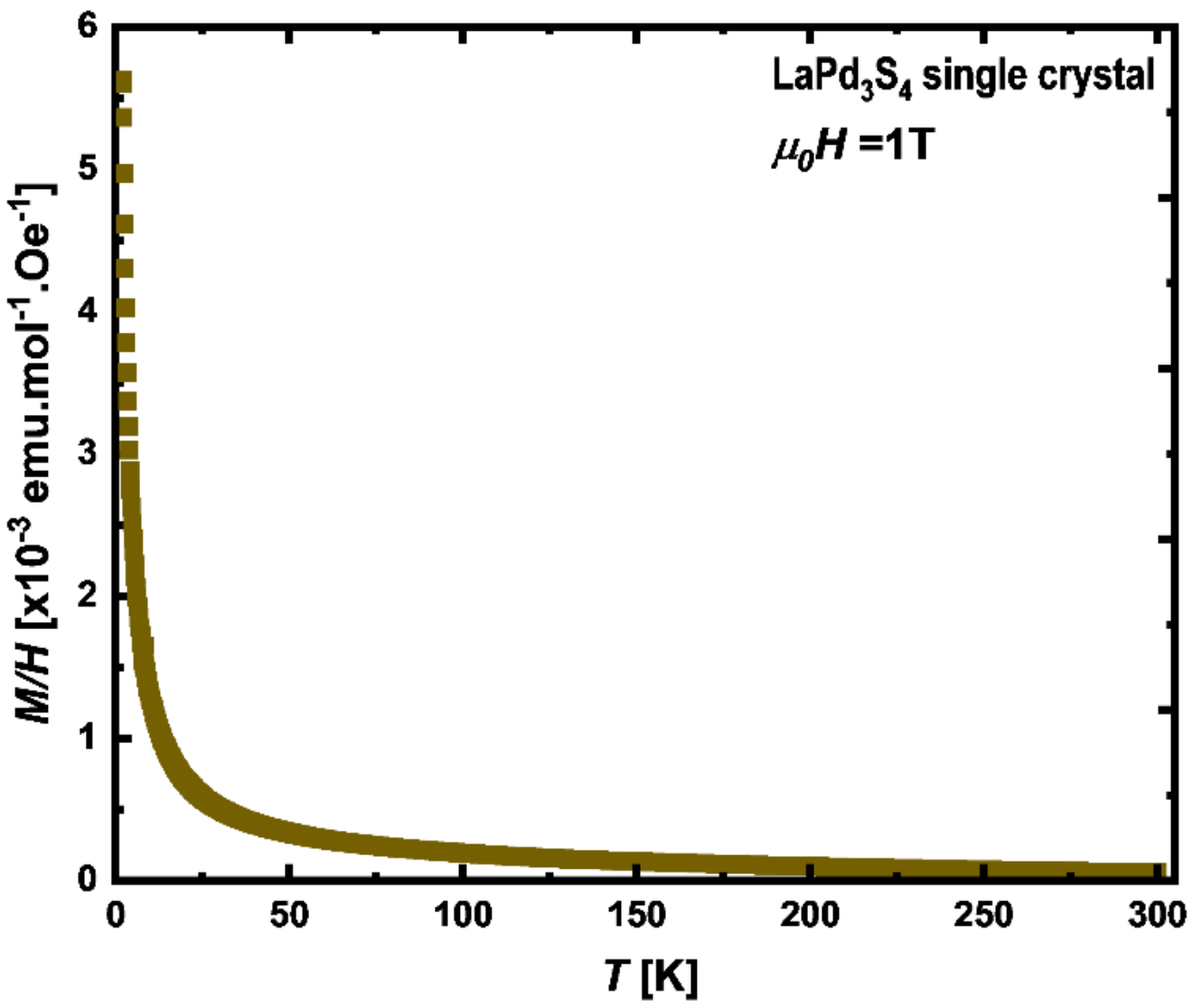

**SI Figure-1** Magnetization as a function of temperature in $LaPd_3S_4$ single crystals at $\mu_0 H$=1 T from $T$≈2–300 K.